\documentclass[12pt,preprint]{emulateapj}

\newcommand{\ltsima}{$\buildrel<\over\sim$}
\newcommand{\lapprox}{\lower.5ex\hbox{\ltsima}}
\newcommand{\msun}{M$_{\odot}$}
\newcommand{\kband}{{\it K$_s$}-band}

\shorttitle{Did New Spheroidals Arise Via Major Mergers?}
\shortauthors{Bundy et al.}
\slugcomment{accepted to ApJL}

\begin{document}

\title{The Mass Assembly History of Spheroidal Galaxies: Did 
 Newly-Formed Systems Arise Via Major Mergers?}

\author{Kevin Bundy\altaffilmark{1}, Tommaso Treu\altaffilmark{2},
  Richard S. Ellis\altaffilmark{3}}


\altaffiltext{1}{Reinhardt Fellow, Dept.~of Astronomy and Astrophysics,
  University of Toronto, 50 St.~George Street, Rm 101, Toronto, ON M5S
  3H4, Canada}

\altaffiltext{2}{Alfred P. Sloan Research Fellow, Dept.~of Physics, University of California,
Santa Barbara, CA 93106-9530}

\altaffiltext{3}{105--24 Caltech, 1201 E. California Blvd., Pasadena,
  CA 91125}

\begin{abstract}

  We examine the properties of a morphologically-selected sample of $0.4
  < z < 1.0$ spheroidal galaxies in the GOODS fields in order to
  ascertain whether their increase in abundance with time arises
  primarily from major mergers. To address this question we determine scaling
  relations between the dynamical mass $M_{dyn}$, determined from
  stellar velocity dispersions, and the stellar mass $M_*$, determined
  from optical and infrared photometry.  We exploit these relations
  across the larger sample for which we have stellar masses in order to
  construct the first statistically robust estimate of the evolving {\em
    dynamical mass function} (MF) over $0<z<1$. The trends observed
  match those seen in the stellar mass functions of \citet{bundy05}
  regarding the top-down growth in the abundance of spheroidal
  galaxies. By referencing our dynamical masses to the halo
  virial mass, $M_{vir}$, we compare the growth rate in the abundance of
  spheroidals to that predicted by the assembly of dark matter
  halos. Our comparisons demonstrate that major mergers do not fully account
  for the appearance of new spheroidals since $z \sim 1$ and that
  additional mechanisms, such as morphological transformations, are
  required to drive the observed evolution.



\end{abstract}


\section{Introduction}

Substantial progress has been made in characterizing the
primarily old, red, and passively-evolving stellar populations of spheroidal
galaxies (defined here to include elliptical and lenticulars) from z$\simeq$1 
to the present day \citep[see][for a review]{renzini06}.  Their evolving 
abundance contains valuable clues to the processes by which they form,
as well as insight into the history of cosmic mass assembly.  In recent 
years, many surveys have converged on a broad evolutionary pattern 
in which it is claimed that galaxies with the largest stellar masses evolve  
predominantly into spheroidals by $z \gtrsim 1$ with little subsequent
growth. Meanwhile, intermediate and lower mass systems continue to form 
at later epochs  \citep[e.g.,][]{bundy05, bundy06, tanaka05, franceschini06,
borch06, pannella06, cimatti06, abraham07}.
  
Some authors \citep[e.g.,][]{van-dokkum05a} have questioned the absence
of growth among massive galaxies, pointing to a high rate of tidal
features in local examples, indicative of recent merging. It is likewise
argued \citep{drory04, van-der-wel06a, maraston06, kannappan07} that
stellar masses derived from optical-infrared photometry may suffer
biases that lead to large uncertainties in the inferred growth rate.
Beyond such concerns, interpreting the observed evolution in the context
of the $\Lambda$CDM paradigm necessitates appealing to potentially
uncertain semi-analytic models.  Clearly what is needed is a way to
connect observations of galaxy growth directly to the assembly history
of the halos in which they reside.

Spheroidal galaxies offer particular advantages in this regard.  First,
because of their regular mass profiles, spectroscopic observations can
be used to infer the dynamical mass, bypassing luminous quantities.
Second, because of their pressure-supported dynamical configurations,
they are expected to result, for the most part, from mergers
\citep[e.g.,][]{toomre77, springel05b}. Accordingly, their evolving
abundances can be compared directly to the expected assembly history of
dark matter halos.  In this Letter we develop a technique to estimate
dynamical and halo masses for a large sample of field spheroidals and
use it to interpret their evolution in the context of halo assembly as
predicted in numerical simulations.  We assume $\Omega_{\rm M}=0.3$,
$\Omega_\Lambda=0.7$, $H_0=70 h_{70}$ km~s$^{-1}$~Mpc$^{-1}$.

\section{Observations and Mass Estimates}\label{data}

The dynamical mass (hereafter $M_{dyn}$) measurements used in this
Letter are based on the analysis presented in \citet{treu05a}, to which
we refer for details. Briefly, this sample comprises 165 field
visually-classified spheroidals selected from ACS imaging in the
northern GOODS field to a magnitude limit of $z_{AB} < 22.5$ and studied
with the DEIMOS spectrograph on Keck II.  A subset of 125 galaxies have
reliable stellar velocity dispersions, stellar masses, and surface
photometry fits, which provide $B$-band effective radii, $R_{eB}$
\citep[see][]{treu05a}.  Dynamical masses were derived using the
formula, $M_{dyn} = K_v \sigma^2 R_{eB}/G$ where $K_v$ is the virial
coefficient and $\sigma$ is the {\it central} velocity dispersion,
obtained by increasing the observed dispersion by 10\% to correct to the
standard circular aperture of diameter $R_{\rm eB}/4$. Detailed
modelling of fiber as well as 3D spectroscopy of nearby ellipticals
supports the use of this equation \citep{padmanabhan04, cappellari06}
with little uncertainties due to the overall flatness of the velocity
dispersion profile.

With the kinematic sample defined in this way, the investigation that
follows is based on the much larger sample of galaxies with stellar
masses (hereafter $M_*$) in both GOODS fields with $z_{AB} < 22.5$
\citep{bundy05}. We refer to that paper for further details but
summarize the key features.  For $0.4 < z < 1.0$, spectroscopic
redshifts for 633 galaxies from the Keck Team Redshift Survey
\citep[KTRS,][]{wirth04} were supplemented with 695 determined using
photometric techniques \citep[see][]{bundy05}.  The full sample of
morphologically-selected spheroidals includes 393 sources. Using deep
\kband\ imaging obtained at Palomar Observatory for GOODS-N and publicly
available data in GOODS-S, stellar masses were estimated assuming a
Chabrier \citep{chabrier03} initial mass function (IMF) using the code
described in \citet{bundy06}\footnote{The results can be adjusted to
  those for a Salpeter IMF by adding $\sim$0.25 dex to the derived
  masses.}.  For galaxies with photometric redshifts, an additional 0.3
dex is added in quadrature to the base uncertainty which is typically
0.2 dex \citep{bundy05,bundy06}.

As discussed in \citet{bundy05}, the GOODS morphological sample is
complete in the \kband\ but becomes incomplete for very red galaxies
below a certain stellar mass limit as a result of the $z_{AB} < 22.5$
limit (see their Figure 4).  By considering synthetic SEDs representing
red (in $z - K_s$ color) stellar populations, we determine mass limits
of $\log (M_{limit}/M_{\odot}) = [9.8, 10.4, 10.8]$ for redshifts, $z =
[0.4, 0.7, 1.0]$.  The completeness limits of the MFs below are
determined at the near edge of each redshift interval.  When comparing
$M_*$ to $M_{dyn}$ we employ the more strict, volume-limited limits at
the far edge of each interval \citep[see discussion in][]{bundy06}.

\section{Linking Dynamical Masses and Stellar Masses}

We begin by restricting our analysis to those 125 spheroidals in
GOODS-N for which both $M_{dyn}$ and $M_*$ measurements are
available. The aim is to develop scaling relations that can be applied
to the larger GOODS-N/S sample. For local galaxies, a fairly tight
correlation of the form $M_{dyn} = (M_*)^{\alpha}$ has been observed
by several authors \citep[e.g.,][]{padmanabhan04, cappellari06,
gallazzi06}.  \citet{gallazzi06} exploited a local sample of
early-type systems selected by concentration, and find $\alpha = 1.28
\pm 0.03$---the zeropoint is not explicitly solved for.  In this work,
we adopt a zeropoint such that $M_{dyn} \equiv M_*$ at $10^{11}$\msun,
and also apply it to the \citet{gallazzi06} relation.  This choice
corresponds to $K_v=4.8$ and 6.6 at $0.4 < z < 0.7$ and $0.7 < z <
1.0$ and is motivated by the fact that the $K_v$ depends on the mass
density profile and, most importantly, on the spectroscopic aperture
used to measure $\sigma$. Since the aperture is effectively
redshift-dependent (the slit width is fixed at 1\arcsec), assuming a
constant $K_v$ could introduce a spurious redshift dependency.
Similarly, while there is no evidence that the density profile evolves
for high mass spheroidals \citep[e.g.,][]{koopmans06}, potential
evolution could mimic changes in the MF if $K_v$ were held fixed. More
detailed studies are needed in order to determine the normalization
independently.

\begin{figure}
\plotone{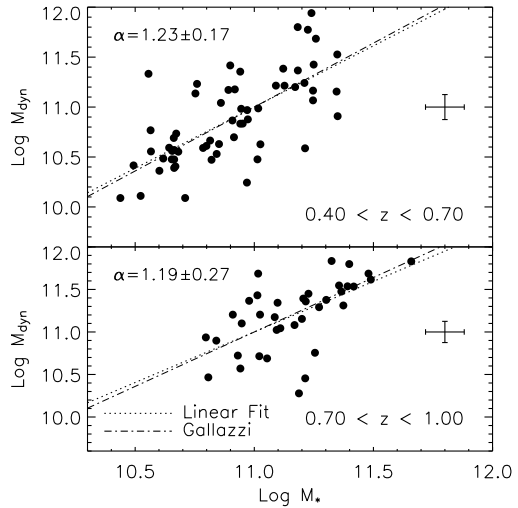}
\caption{Correlation between $M_{dyn}$ and $M_*$.  The dotted line
  illustrates the fitted slope, $\alpha$, which is also indicated in the
  top left corner.  A typical error bar is shown and the relation found
  by \citet{gallazzi06} is also plotted.  The normalization in each
  panel has been set to the same scale, as described in the
  text.  \label{relplot}}
\end{figure}

We examine the $M_{dyn}$--$M_*$ relation at $0.4 < z < 0.7$ and $0.7 <
z < 1.0$ in Figure \ref{relplot}, restricting the comparison to
systems above our completeness limit; the best fit values for $\alpha$
are also indicated. Our observed slopes are formally consistent with
those of \citet[$1.28\pm0.03$]{gallazzi06} at $z=0$ and
\citet[$1.25\pm0.05$]{Rettura06} at $z \sim 1$ \citep[see also]{di-serego-alighieri05}, although some
differences may be expected as a result of sample selection and fiber-
versus slit-based measures of velocity dispersion.  The fact that
$\alpha > 1$ reflects the ``tilt'' of the Fundamental Plane
\citep[e.g.][]{faber87,CLR96,renzini06} and is consistent with
non-homology \citep[e.g.][]{TBB04} including an increasing dark matter
fraction within $R_e$ at higher masses \citep[e.g.][]{padmanabhan04,
gallazzi06, bolton07}.  With regard to what follows, the key result in
Figure \ref{relplot} is that $\alpha$ shows no measurable evolution
across our sample.  In $\S$\ref{section:DMF}, we will show that this
leads to dynamical MFs that behave similarly to their stellar mass
counterparts.  For consistency, final $M_{dyn}$ estimates for galaxies
in the full GOODS sample are determined using the value of $\alpha$
determined for the two redshift bins, and the final uncertainty in
$M_{dyn}$ is taken to be the sum in quadrature of the uncertainty in
$M_*$ and the scatter in Figure \ref{relplot}.

\begin{figure}
\plotone{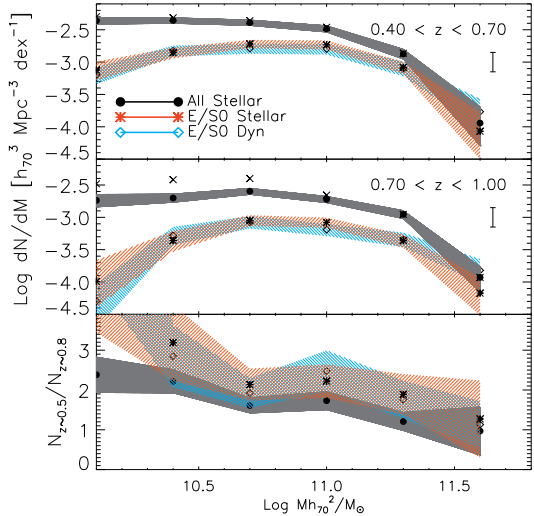}
\caption{Comparison of evolving dynamical and stellar mass functions.
  The top two panels show the MFs in two redshift intervals, while the
  bottom panel illustrates the differential growth between them.
  Solid points with gray shading trace the stellar mass function of
  the full (all types) GOODS $z_{AB} < 22.5$ sample, while crosses
  denote the equivalent $z_{AB} < 23.5$ sample.  Asterisks with light red shading
  trace the stellar MF of the spheroidal component, while diamonds
  with blue shading indicate the dynamical MF.  The fact
  that the observed increase with time in the 
  spheroidal MFs is more significant than the increase observed for the full
  sample (bottom panel), shows that this result is robust to potential
  incompleteness.  Isolated error bars
  indicate the estimated uncertainty from cosmic variance.
  \label{lvm_comp}}
\end{figure}

\section{The Dynamical Mass Function}\label{section:DMF}

Using the $M_{dyn}$ estimates derived above for the full GOODS sample,
we now construct the dynamical MF and compare with its
stellar mass counterpart in the top two panels of Figure \ref{lvm_comp}.
We use the $1/V_{max}$ estimator and closely follow the method described
in \citet{bundy05}. The solid points with grey shading show the stellar MF for
the full sample, while the crosses indicate the full stellar MF that
results when the magnitude limit is relaxed to $z_{AB} < 23.5$.  The
bottom panel of Figure \ref{lvm_comp} shows the differential growth
derived across the two redshift bins.


Figure \ref{lvm_comp} demonstrates consistency between the stellar MFs
and dynamical MFs. Although our adopted zeropoint in the $M_{dyn}$
relations ensures a similar vertical normalization, the application of
our scaling relations does not mean both MFs should have the same shape.
Indeed, small differences are noticeable in our highest mass bin.  The
good agreement results in large part from the fact that the
$M_{dyn}$--$M_*$ relation does not vary significantly across our sample.
Had a variation in $\alpha$ with redshift been present in Figure
\ref{relplot}, the resulting dynamical MFs would evolve with respect to
the stellar MFs.  The fact the two agree validates earlier stellar mass
estimates at $z \sim 1$ and the top-down growth in the abundance of
spheroidals derived from those measures.


\section{Testing the Role of Mergers}

We now turn to the key question which our analysis can address: is the
rising abundance with time of spheroidal galaxies consistent with that
predicted by the growth of merging dark matter halos?  To answer this
question we adopt a simple, observationally-motivated model to connect
$\sigma$ and $M_*$ to the virial mass, $M_{vir}$, of the host halos.
Lensing studies of early-type galaxies at $z \approx 0.2$
\citet[see][and references therein]{gavazzi07} show that the total mass
density profile of spheroidal halos is nearly isothermal within the halo
scale radius, $r_s$.  We therefore assume a mass profile that is
isothermal for $r \lesssim r_s$ and follows a $\rho \propto r^{-3}$
profile \citep[NFW,][]{navarro97} for $r \gtrsim r_s$.  We note that the
$\rho \propto r^{-3}$ behavior at large radii ($r \sim r_{vir}$) is consistent with non-NFW
scaling laws which can disagree instead at small radii
\citep[e.g.,][]{graham06}.  The mass profile assumed here is given by
$\rho(r) = \sigma^2 [2 \pi G r^2 (1+r/\gamma r_s)]^{-1}$, where
$\sigma$ is the observed central velocity dispersion and we have
introduced the scaling parameter, $\gamma$, which can be used to tune
the profile shape to observations.  By comparing to \citet{gavazzi07},
we find $\gamma = 12$---future observations will constrain $\gamma$ as
a function of redshift and mass.  We define $M_{vir} = \frac{4\pi}{3}
\Delta(z) \rho_c c^3 r_s^3$, where $\rho_c$ is the cosmic critical
density, $\Delta(z)$ is the overdensity parameter
\citep[see][]{bryan98}, and $r_{vir}= c r_s$.  The concentration, $c$, is assumed to the
follow the relation found in simulations, $c = 9(1+z)^{-1}(M_{vir}/(8.12
\times 10^{12}h^{-1} M_{\odot}))^{-0.14}$ \citep{bullock01}.  Using this
model and the relations above, we solve for $M_{vir}$ as a function of
velocity dispersion and find,

\begin{equation}
\log M_{vir} = A_{\sigma} \log \sigma_{200} + B_{\sigma} + C_{\sigma}(1+z),
\label{eqn_sigma}
\end{equation}

\noindent where $M_{vir}$ has units of $h_{70}^{-1}$\msun,
$\sigma_{200}$ is the velocity dispersion in units of 200 km s$^{-1}$,
and the fit parameters are given by $A_{\sigma}=3.1$,
$B_{\sigma}=13.2$ and $C_{\sigma}=-0.3$.  Equation \ref{eqn_sigma}
follows from the model described above and so random uncertainties on
the fitting parameters are negligible.  However, systematic errors
resulting from tuning this model to observations should reflect the
uncertainty found by \citet{gavazzi07} of 0.2 dex.  Using our
kinematic sample for which we measure both $\sigma$ and $M_*$, we can
also directly compare $M_{vir}$ and $M_*$, finding the following
relation:

\begin{equation}
\log M_{vir} = A_{*} \log M_{*11} + B_{*} + C_* (1+z)
\label{eqn_mstar}
\end{equation}

\noindent where a Chabrier IMF was assumed as before and $M_{*11}$ has
units of $10^{11} h_{70}^{-2}$\msun.  The best fitting parameters are
$A_{*}=1.0 \pm 0.2$ and $B_{*}=12.83 \pm 0.11 \pm \sigma_{B,sys}$,
$C_*=-0.2 \pm 0.1$ where $\sigma_{B,sys} = 0.25$ dex and accounts for
the systematic uncertainty in Equation \ref{eqn_sigma} (0.2 dex) as well
as uncertainty in the stellar IMF (0.15 dex). Equations 1 and 2 are
applicable for galaxies with $\log
M_*/M_\odot \approx 10-12$ and $\sigma \approx 90-300$ kms$^{-1}$.  Equation
\ref{eqn_mstar} gives values of $M_{vir}/M_* = 30^{+25}_{-15}$,
consistent with \citet{mandelbaum05} and \citet{gavazzi07}.  We note
that our marginally non-zero value of $C_*$, if confirmed, implies a
slight increase with time of $M_{vir}/M_*$ ($\sim$30\%) over $0.4 < z
< 1.0$ that is consistent with halo growth \citep[see discussion
in][]{conroy07} and independent evidence that stellar mass assembly is
mostly completed by $z \approx 1$ \citep[e.g.][]{treu05a}. We also
note that virial mass and stellar mass turn out to be approximately in
linear proportion within this mass range.  Studies with a larger
dynamical range in stellar masses and velocity dispersion, and more
weak lensing measurements, are needed to measure the slope with higher
precision and detect departures from linearity, expected as a result
of varying star formation efficiency with halo mass.

\begin{figure}
\plotone{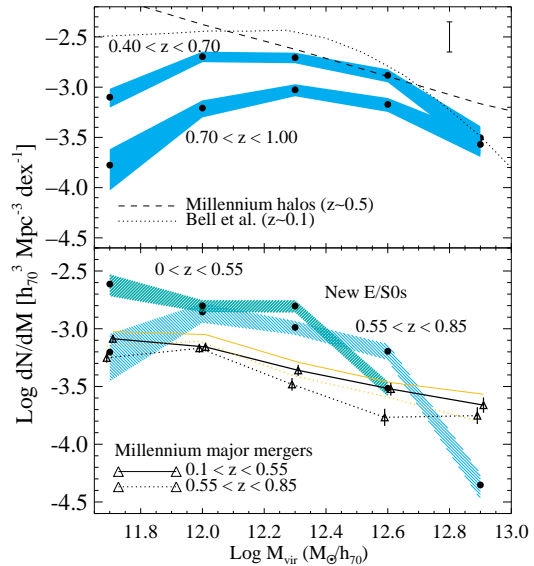}
\caption{Evolving virial mass functions of halos hosting spheroidals.
  The top panel shows MFs in two $z$-bins. The local MF estimated from
  \citet{bell03}, is indicated by the dotted line.  The dashed line
  shows the total MF of predicted dark matter halos, and the isolated
  error bar (top right) denotes the cosmic variance uncertainty.  The
  bottom panel shows the MF of newly formed spheroidals determined by
  subtracting the MFs (top panel) in adjacent redshift bins.
  Overplotted are two halo merger scenarios based on the Millennium
  Simulation, integrated over the redshift ranges indicated.  Black
  lines with triangle symbols trace $>$1:3 mergers while yellow lines
  (without symbols) denote such mergers with a $\sim$0.5 Gyr time
  delay.  \label{mvm_evol}}
\end{figure}

Using Equation \ref{eqn_mstar} we estimate halo virial masses for our sample and compare the resulting virial MFs in the
top panel of Figure \ref{mvm_evol}.  The dashed line is the total halo
mass function predicted at $z \approx 0.5$ (it evolves little over
$0<z<1.0$) using the online Millennium Simulation
Database\footnote{http://www.mpa-garching.mpg.de/millennium.  The
  ``milli-Millennium'' Database covers a volume of $7 \times
  10^{5}$~$h_{70}^{-3}$Mpc$^3$, nearly 4 times larger than our largest
  sampled volume at $0.7 < z < 0.4$.} \citep{springel05a} and adjusted
in number density by $+$0.1 dex to be consistent with the spheroidal MFs
(this adjustment is within the expected cosmic variance
of 0.15 dex).  The dotted line shows a local ($z \approx 0.1$) estimate constructed by
converting the $M_*$ MF for early-types (based on concentration, $C \ge
2.6$) measured by \citet{bell03} into a virial MF using Equation
\ref{eqn_mstar}.  Adjustments of $+$0.1 dex were also applied to the
local number density and the local mass scale to match the GOODS sample
at high masses.  Such corrections are likely needed as a result of
cosmic variance, differing selection and mass estimate methods, and
photometric uncertainties.  Thus, comparisons between
our results and the local MF must be done with caution.

The top panel of Figure \ref{mvm_evol} reveals a similar pattern of
evolution as seen in Figure \ref{lvm_comp}.  A useful interpretation of
this evolution is given in the bottom panel of Figure \ref{mvm_evol}
which plots the $\log$ {\em difference} ($\log \Delta dN/dM$), thereby
displaying the halo MF of new spheroidals that have appeared over the
time interval separating the two redshift windows and providing a way to
compare with the incidence of halo mergers.  We use the MPAHalo
milli-Millennium database to identify halos that have experienced a
recent merger, at which point two previously separate halos or subhalos
become a single halo.  Halo merging often occurs several Gyr after halo
accretion, defined by the point at which the two progenitor halos first
become associated with the same friends-of-friends (FOF) group.  Because the
smaller progenitor halo may experience significant tidal stripping after
halo accretion but before merging \citep[see][]{gao04}, we take the
merger mass ratio to be the initial ratio before halo accretion.

After integrating over the two time intervals that separate our
observations, we overplot two merging scenarios in the bottom panel of
Figure \ref{mvm_evol} (the $+$0.1 dex vertical offset is also applied).
It is a striking result that major halo mergers (mass ratios greater
than 1:3, indicated by dark lines) cannot account for the rise of
spheroidal systems over the two epochs probed here.  The light yellow
lines (without symbols) show this is still the case when a $\sim$0.5 Gyr
delay is added to the merger timescale (as suggested by expected
differences in the galaxy and halo merger times).  We note that at the
highest masses probed there is an indication that the predicted merger
frequency rises above the growth of new spheroidals.  This might be
expected since most systems at such masses already exhibit spheroidal
morphologies by these redshifts \citep[e.g.,][]{bundy05}, suggesting
such interactions represent ``dry'' mergers
\citep[e.g.,][]{van-dokkum05a}.

\section{Discussion}

By constructing dynamical mass functions for 393 field spheroidals and
linking those mass estimates to the halos in which they reside, we have
found a significant result in the bottom panel of Figure \ref{mvm_evol}:
major merging of dark matter halos as described by $\Lambda$CDM does not
occur frequently enough to explain the observed increase in the mass
function of spheroidal galaxies.  A check made by appealing to the
semi-analytic model of \citet{de-lucia07} shows a similar result when a
galaxy's stellar mass is used to define recent mergers and even when
$M_*$ is used as the ``accounting variable.'' 

If merging is not the only mechanism that produces newly-formed
spheroidals since $z \sim 1$, other physical processes likely play a
significant role.  These may include mechanisms for
transforming disk and irregular galaxies into relaxed spheroidals, for
example secular bulge growth accompanied by disk fading.  Indeed, recent
semi-analytic models \citep[e.g.,][]{croton06, bower06, de-lucia07} have
required prescriptions for such mechanisms, in addition to mergers, to match the local
abundance of spheroidals.  More work is needed to understand these
processes and their role in morphological evolution.

\acknowledgments 

We are very grateful to Simon White and Volker Springel for help
in deriving results from the Millennium Run Database and also thank Raphael
Gavazzi, Andrew Benson, Phil Hopkins, and Karl Glazebrook for useful
discussions.  RSE acknowledges the hospitality of Ray Carlberg at the
University of Toronto during a sabbatical visit and TT acknowledges
support from a Sloan Research Fellowship.


\end{document}